\documentclass[aps,preprint,amssymb,12pt,floatfix]{revtex4}
\usepackage{subfigure}
\usepackage{graphicx}
\usepackage{rotating} 
\usepackage{color}
 \usepackage{amsmath,amsthm}
\usepackage{epstopdf}

\begin{document}
%\draft

\title{Membrane-protein interactions hold the key to understanding amyloid formation}

\author{John E. Straub$^{1}$ and D. Thirumalai$^{2}$}

\affiliation{
$^1$Department of Chemistry, Boston University, Boston, MA, 02215\\
$^2$Department of Chemistry and Biochemistry, and Biophysics Program, Institute for Physical Sciences and Technology, University of Maryland, College Park, MD 20742}
\date{\today}

\maketitle  

Spurred by the appreciation that protein aggregation, leading to amyloid fibril formation, is linked to a growing list of diseases there has been intense effort to understand the factors that trigger association between peptides. This vastly complex field has attracted a variety of approaches ranging from mutational studies at the genetic level to characterization of events at the molecular level of specific proteins. It is unclear if biophysical approaches used to study protein folding in quantitative terms will have a major impact on leading to a cure or management of any amyloid disease, including Alzheimer's Disease (AD). Nevertheless, it is clear that rigorous biophysical methods are needed to gain a deeper understanding of the routes to amyloid formation. After nearly fifteen years of {\it in vitro} experimentation and simulations, certain general principles that govern oligomer (suspected to be the cause of toxicity) and amyloid formation have emerged. A few of the most important findings are: (1) All proteins, regardless of whether their aggregation is related to any specific disease, can form amyloid fibrils with characteristic cross $\beta$ structure \cite{Chiti06ARBiochem}. This finding implies that under suitable conditions favoring aggregation  the folded monomeric state is unlikely to be the most stable.  Although the possibility that folded proteins are metastable was suggested long ago \cite{Honeycutt90PNAS}, only recently it has been argued that under nominal protein concentrations found in cells the native state is metastable with respect to the fibril state \cite{Baldwin11JACS}. (2) The propensity of a polypeptide chain to aggregate is encoded in the the free energies of the spectrum of states of the monomer \cite{Tarus06JACS}. There is a direct correlation between the probability of accessing such aggregation-prone states and the rate of amyloid formation \cite{Li10PRL}. (3) Although there are exceptions, solid state NMR experiments \cite{Tycko13ACR} on a number of systems, show that in the amyloid fibril state the polypeptide chains are arranged as parallel $\beta$-sheets, which given the conclusion summarized in (1) implies that the ground state of interacting peptides has universal structural features. (4) Despite the possible universal structure of the fibrils,  precise sequence plays a major role in the rate of protein aggregation as well as in the stability of the final product \cite{Goldschmidt10PNAS}. Dramatic variations in aggregation rates of mutants of A$\beta$ peptides \cite{Yamamoto04JNeurochem}, implicated in AD, validates the importance of sequence.

The general principles that have been discovered, using well-defined systems summarized above are of great significance, and establish the power of biophysical methods. However, the studies leading to these conclusions do not even approximately model the process of {\it in vivo} aggregation. For example, it is well-known that in order to understand amyloid formation {\it in vivo} it is important to consider interactions of aggregating proteins with membranes, which in the biophysics community has received little attention. In the interesting Perspective appearing in this issue Bucciantini et. al. \cite{Bucciantini14JPCL} review the specific roles that membranes play in promoting oligomer formation in amyloidogenic proteins with special emphasis on A$\beta$ peptides. The authors have managed to navigate a substantial portion of the still growing literature to summarize the key membrane characteristics, which are involved in promoting or suppressing oligomer formation. Based on general considerations it is clear that membrane-mediated aggregation should depend on physical properties, such as net charge per unit area and hydrophobicity and hydrophilicity. Indeed, it has been found that membranes composed of anionic phospholipids (such as phosphatidylserine and phosphatidyglycerol) catalyze the formation of amyloids \cite{Necula03JBC}.  The mechanical properties of the fluid membranes also affect interaction between proteins, as demonstrated using continuum theories. Thus, cholesterol (Ch) or crowding agents that effect the overall membrane fluidity also affects aggregation \cite{Kim05JBC}, as summarized succinctly by   \cite{Bucciantini14JPCL}. It is likely that theoretical ideas in the soft matter field could be profitably used to predict generic aspects of membrane-mediated interpeptide interaction. 

The specific details of membrane composition, which cannot be easily taken into account using using only the general physicochemical properties, requires a deeper understanding of biological membranes. For example, biological membranes form lipid rafts, which are best pictured as heterogeneous structures. In rafts in neuronal cells, GM1 (see Fig. 2 in \cite{Bucciantini14JPCL}) is the most abundant ganglioside, and is implicated in AD, Parkinson, and Huntington diseases. Interaction of GM1 with A$\beta$ might be involved in promoting A$\beta$ aggregation. Further complicating matters, Ch, whose role in affecting protein aggregation is not fully understood, appears to encourage the formation of islands that are rich in GM1, reminiscent of microphase separation. Based on a study of several model systems \cite{Tashima04JBC}, it has been asserted that in a lipid mixture composition resembling that found in the cerebral cortex, A$\beta$ peptides were rapidly released resulting in amyloid fibril formation. However, in membranes devoid of Ch and GM1, aggregation did not occur, suggesting that both Ch and GM1 control A$\beta$ aggregation in neuronal cells.  In an insightful study, it has been shown that the Ch effect depends roughly on $\lambda$ = [Ch]/[Pl], where [Ch]([PL]) is the concentration of Ch (phospholipids). At low $\lambda$ values A$\beta$ aggregation is promoted whereas at higher values there is increased insertion of A$\beta$ into the membranes, which suppresses aggregation. Therefore, it is important to examine the phase diagram of the ternary (minimum) of Ch, lipids, and A$\beta$ before any general conclusion can be drawn. It is worth emphasizing that lipid composition, which greatly complicates the picture, is an important variable. Because of this model membrane systems are extremely useful in obtaining insights into the aggregation process. 

A critical factor that is receiving increasing attention is the determination of the  structures and dynamics of the type I transmembrane Amyloid Precursor Protein (APP) found in neural and non-neural cells.   The cleavage of APP resulting in
A$\beta$-peptide of varying lengths is achieved by secretases \cite{Wolfe07JCellSci}. Cleavage occurs in several steps. $\beta$-secretase cleaves APP at the $\beta$-site, and the extracellular
domain of APP disassociates from the remaining protein (APP-C99). Subsequently, $\gamma$-secretase  cleaves processively with normal termination at
the $\gamma$-site, which is located on the transmembrane domain of APP-C99, and the product A$\beta$-protein is
released to the extracellular region (Fig. 1).   Many of the familial AD-associated mutations occur close to the secretase cleavage sites
(Fig. 1) and, intriguingly, their is evidence that mutations intermediate to the $\beta$- and $\gamma$-sites may also influence the A$\beta$ product distribution.  A major focus
of current research on AD is to understand the elementary steps in the  cleavage process, with the hope that the design of drugs that modulate the production of deleterious A$\beta$ products requires a molecular description of the structures of APP-C99 in the presence of membranes. Simulations as well as experiments have firmly established that the flexibility of the GG kink \cite{Miyashita09JACS,Barret12Science} near the $\gamma$-site plays an important role in the production of distinct isoforms of A$\beta$ upon cleavage of APP-C99 by $\gamma$-secretase. It is clear that a global understanding of AD (and related diseases) can only be achieved through knowledge of the molecular structures of APP-C99, the details of the cleavage processes by secretases, and characterization of subsequent events leading to aggregation. All of these events involved in AD depend on lipid composition, Ch, and other factors,  thus making the biophysics of AD an extraordinarily rich field, which is sure to attract physics-based approaches for years to come. 

\newpage
%\bibliography{Agg}

\newpage
\begin{figure}{}
\begin{center}
\centerline{\includegraphics[width=.95\textwidth]{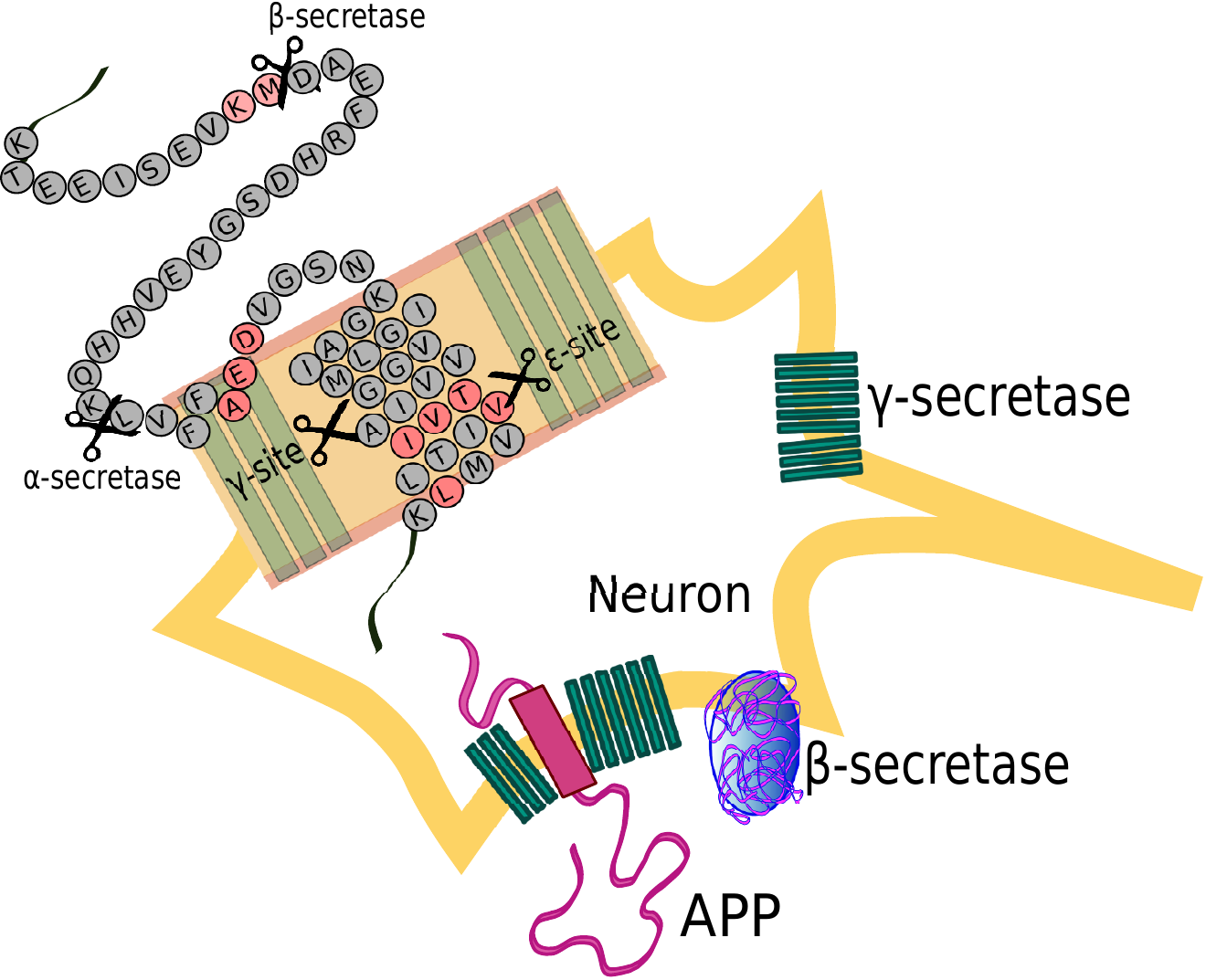}}
\caption{Schematic picture of the Amyloid Precursor Protein in a membrane. On the left the sequence of a portion of APP is shown along with the sites for engaging the secretases. Cleavage by the flexible GG kink in the inserted portion of APP near the $\gamma$-site produces A$\beta$ peptides with 39 - 43 amino acid residues. On the right a picture of the neuron and the way that APP interacts with the membrane of the neuron cell is shown along with the different secretases.
}
\end{center}
\end{figure}

\end{document}